\documentstyle[twocolumn,aps,epsf,floats]{revtex}
\begin{document}
\draft

\twocolumn[\hsize\textwidth\columnwidth\hsize\csname@twocolumnfalse\endcsname

\title {Anisotropic magnetoresistance of GaAs two-dimensional holes}
\author{S. J. Papadakis, E. P. De Poortere, and M. Shayegan}
\address{Department of Electrical Engineering, Princeton University,
Princeton, New Jersey  08544, USA.}
\author{R. Winkler}
\address{Institut f\"ur Technische Physik III, Universit\"at
Erlangen-N\"urnberg, Staudtstr. 7, D-91058 Erlangen, Germany.}
\date{\today}
\maketitle
\begin{abstract}
Experiments on high-quality GaAs (311)A two-dimensional holes at
low temperatures reveal a remarkable dependence of the
magnetoresistance, measured with an in-plane magnetic field ($B$),
on the direction of $B$ relative to both the crystal axes and the
current direction.  The magnetoresistance features, and in
particular the value of $B$ above which the resistivity exhibits
an insulating behavior, depend on the orientation of $B$. To
explain the data, the anisotropic band structure of the holes and
a re-population of the spin-subbands in the presence of $B$, as
well as the coupling of the orbital motion to $B$, need to be
taken into account.
\end{abstract}

\vskip1pc]

There has been interest recently in the ground state of a
disordered two-dimensional (2D) carrier system.  Twenty years ago,
scaling arguments and supporting experimental data indicated that
at temperature $T$ = 0 K such a system must be insulating
\cite{Abrahamsscaling,BishopTsui}. However, prompted by new data
in Si 2D electrons \cite{KravMIminus} and subsequently multiple
different 2D carrier systems \cite{MIpapers}, revealing a
metallic-like behavior, this question is being revisited both
experimentally and theoretically
\cite{MItheory,Pudalov97,DasSarma99,DasSarmaC99}.

One specific area of interest has been the spin degree of freedom
\cite{Pudalov97,Murzin98,Papadakis99,Papadakis99c,Yaish99}.
Measurements have shown that increasing the spin-orbit induced
zero-magnetic-field spin-splitting leads to a more pronounced
metallic behavior in GaAs 2D holes
\cite{Papadakis99,Papadakis99c}.  References
\cite{Murzin98,Yaish99} also report that the metallic behavior in
this system is related to transport by two spin-subbands.
Experiments with an in-plane magnetic field ($B$) similarly
suggest that the effects of spin are important
\cite{Pudalov97b,Simonian97,Mertes99,Okamoto99,Yoon99b}. On the
other hand, for 2D systems with finite layer thickness recent
calculations predict an anisotropic positive magnetoresistance
(MR) caused by the coupling of the orbital motion to $B$
\cite{DasSarmaC99,Chen-more}.  The MR is calculated to be larger
for $B \perp I$ than for $B \parallel I$, where $I$ is the current
in the sample. Motivated by this prediction, we measure the MR of
a high-mobility 2D hole system (2DHS) in a GaAs (311)A quantum
well, in a density range such that the $B = 0$ sample resistivity
($\rho$) shows metallic $T$-dependence. We apply an in-plane $B$
parallel to the $[\bar233]$ and $[01\bar1]$ crystal axes, and
measure the MR with $I \parallel B$ and $I \perp B$ for each case.
Some characteristics of our data, such as an overall positive MR,
are consistent with the predictions of Ref. \cite{DasSarmaC99}.
However, we observe a striking dependence of the MR, and in
particular of the {\it onset of insulating behavior}, on the
orientation of $B$ relative to the {\it crystal axes}. We show
that this dependence is linked to the anisotropy of the 2DHS band
structure, and a re-population of the spin-subbands with
increasing $B$.

The samples are Si-modulation doped GaAs quantum wells grown on
(311)A GaAs substrates.  Even at $B = 0$, these samples exhibit a
mobility anisotropy believed to be due to an anisotropic surface
morphology (see \cite{Heremans94,Wassermeier95} and references
therein). The interfaces between the GaAs quantum well and the
AlGaAs barriers are believed to be corrugated, with ridges along
the $[\bar233]$ direction which reduce the mobility for $I
\parallel [01\bar1]$. While the metallic behavior has been studied
in this system extensively, little attention has been paid so far
to the differences between transport along $[01\bar1]$ and
$[\bar233]$. Our sample is patterned with an L-shaped Hall bar
aligned along $[01\bar1]$ and $[\bar233]$ to allow simultaneous
measurement of the resistivities along the two directions. The
sample has metal front and back gates to control both the 2DHS
density ($p$) and the perpendicular electric field ($E_{\perp}$)
applied to the well \cite{Papadakis99,Papadakis99c}.  Measurements
are done in dilution and pumped $^3$He refrigerators with $B$ up
to 16 T. In the $^3$He refrigerator, the sample is mounted on a
single-axis tilting stage that can be rotated {\it in-situ} to
change the plane of the 2DHS from perpendicular to parallel to the
applied $B$.

\begin{figure}[tb]
\vskip-1pc
\epsfxsize=3.25in
\epsfbox{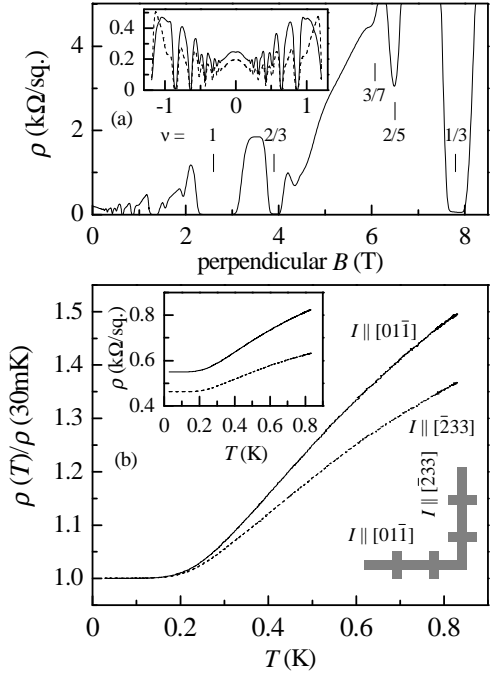}
\vskip-2pc
\caption{(a)  Resistivity $\rho$ data for magnetic field $B$
perpendicular to the plane of the 2D hole system with a density $p
=  6.3 \times 10^{10}$ cm$^{-2}$ and current $I
\parallel [\bar233]$ at $T = 30$ mK.  The data exhibits fractional quantum Hall
effect at low filling factors ($\nu$), demonstrating the high
quality of the sample. The inset shows low-$B$ data for $I
\parallel [01\bar1]$ (solid trace) and $I
\parallel [\bar233]$ (dashed trace). (b) $B = 0$ temperature-dependence
at $p = 3.3 \times 10^{10}$ cm$^{-2}$, highlighting the difference
between $[01\bar1]$ (solid) and $[\bar233]$ (dashed) directions.
The main figure shows the fractional change in $\rho$ as $T$ is
increased, while the inset shows the raw data. The schematic at
lower right depicts the Hall bar used for the measurements.}
\label{aniso}
\end{figure}

Figure \ref{aniso}(a) demonstrates the high quality of the 2DHS in
our sample.
The data of Fig. \ref{aniso} also reveal the mobility anisotropy
observed in this sample: at 30 mK and $p = 6.3 \times 10^{10}$
cm$^{-2}$, we have $\mu_{[01\bar1]} = 425,000$ cm$^2$/Vs and
$\mu_{[\bar233]} = 530,000$ cm$^2$/Vs.  As illustrated in Fig.
\ref{aniso}(b), the $T$-dependence of $\rho$ is also significantly
different along the two directions in the density range where the
behavior is metallic. The $[01\bar1]$ direction typically shows a
larger fractional change in $\rho$, ${\rho}(T)/{\rho}(30$ mK),
than the $[\bar233]$ direction, as $T$ is increased
\cite{Papadakis99,Papadakis99c}. This suggests that the scattering
mechanisms associated with the two mobility directions have
different $T$ dependencies, and that the orientation of $I$
relative to the crystal axes is an important parameter in
understanding the data.

\begin{figure}[tb]
\epsfxsize=3.25in
\epsfbox{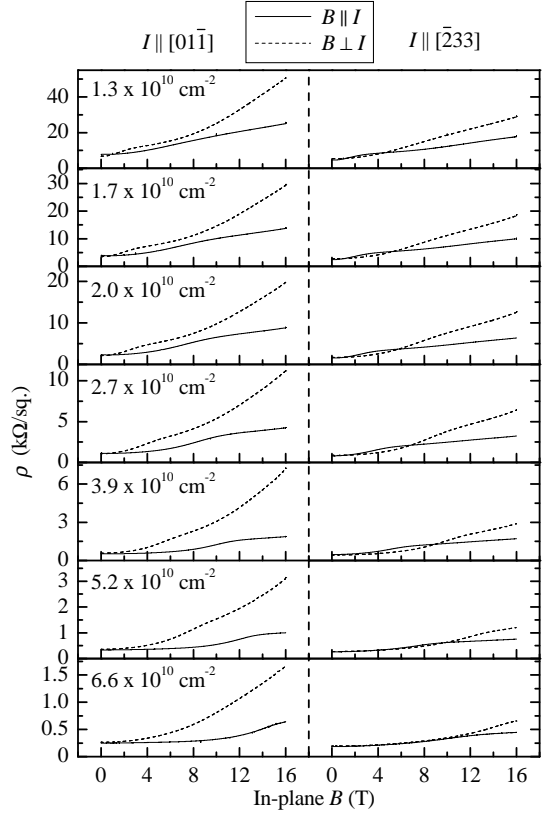}
\caption{Magnetoresistance at $T = 0.3$ K in an in-plane $B$, showing how
resistivity changes as the orientation of $B$ is changed from
parallel to perpendicular to the current ($I$) direction.  The
density $p$ is listed in each panel.} \label{intro}
\end{figure}

Figure \ref{intro} shows $\rho$ at $T = 0.3$ K as a function of
$B$ applied in the plane of the 2DHS.  The left (right) column
shows data for $I \parallel [01\bar1]$ ($I
\parallel [\bar233]$), with the in-plane $B$ both parallel and
perpendicular to $I$.  To obtain these data, on separate cooldowns
the sample was mounted with the $[01\bar1]$ or the $[\bar233]$
crystal axis parallel to the tilt axis.  The density $p$ was
deduced from the Hall coefficient by measuring the transverse MR
in a $B$ perpendicular to the plane of the 2DHS. The stage was
then tilted to make the 2DHS plane parallel to the applied $B$,
and the MR was measured.  The front and back gates were used to
change $p$.  For $I \parallel [01\bar1]$, $\rho$ is always larger
when $I \perp B$.  However, the $I \parallel [\bar233]$ data are
qualitatively different: at low $B$ and $p$, the $I \perp B$
traces have lower $\rho$, and cross the $I \parallel B$ traces at
higher $B$.

\begin{figure}[tb]
\epsfxsize=3.25in
\epsfbox{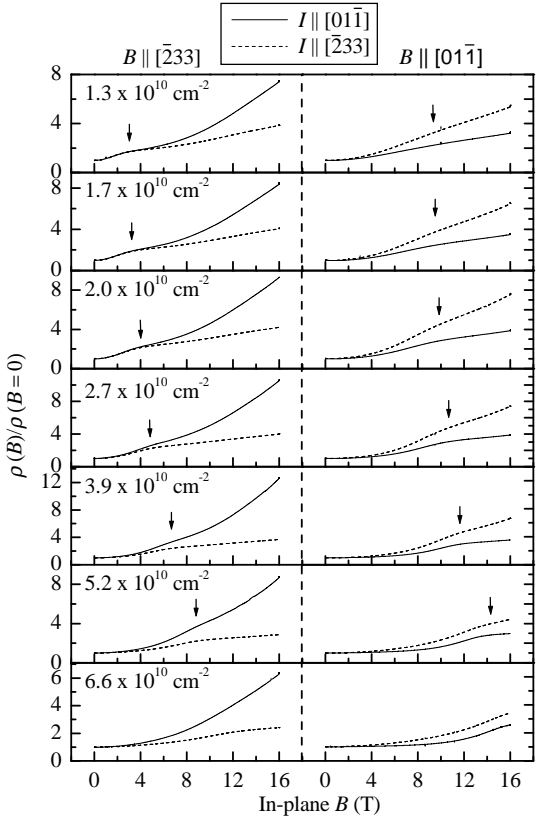}
\caption{Fractional change in resistivity due to an in-plane $B$,
showing that the relative orientations of $B$ and the crystal axes
play an important role in determining the position of the
magnetoresistance features.  The vertical arrows mark $B^*$ as
defined in the text.} \label{density}
\end{figure}

In Fig. \ref{intro}, there are pronounced qualitative similarities
between the dashed traces on the right and the solid traces on the
left, and vice versa.  To highlight these similarities, in Fig.
\ref{density} we show the fractional change in $\rho$,
${\rho}(B)/{\rho}(B = 0)$, for $B$ along the $[\bar233]$ (left
column) and $[01\bar1]$ (right column) directions.  Plotting this
way, a striking similarity is evident in the qualitative features
of the traces with the same $B$ orientation relative to the
crystal axes, even though the $I$ orientations are different. All
traces start with a small slope and curve upwards. Then there is
an inflection point followed by a reduction in slope, followed by
another inflection point beyond which the traces curve upwards
again. To highlight this behavior, the arrows in Fig.
\ref{density} are placed between the two inflection points, at a
$B$ we will refer to as $B^*$. Surprisingly, for each $p$, $B^*$
for the $B \parallel [\bar233]$ traces is about 4 T smaller than
for the $B \parallel [01\bar1]$ traces, regardless of the $I$
direction. Also, $B^*$ becomes smaller as $p$ is reduced. Figure
\ref{density} reveals that the relative orientations of $B$ and
the crystal axes play an important role in the MR features.

\begin{figure}[tb]
\epsfxsize=3.25in
\epsfbox{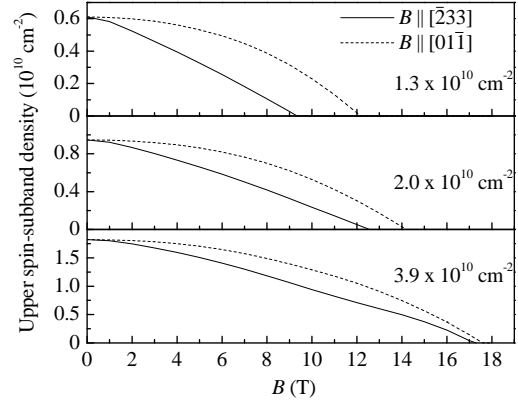}
\caption{Results of self-consistent calculations showing the
upper spin-subband density with increasing in-plane $B$.  Note the
significant difference between curves for $B
\parallel [\bar233]$ and for $B \parallel [01\bar1]$. }
\label{Sim}
\end{figure}

The existence of the MR features around $B^*$ is intriguing.
Similar, though sharper, features have been observed in in-plane
$B$ measurements in systems with multiple confinement subbands
when a subband is de-populated \cite{Jo93}. We propose that the MR
features observed in our data are related to the changes in the
relative populations of the spin-subbands and the resulting
changes in subband mobility and inter-subband scattering as the
in-plane $B$ is increased. To test this hypothesis, we have done
self-consistent subband calculations with no adjustable parameters
\cite{KPtheory} that give us spin-subband densities as a function
of in-plane $B$ (Fig. \ref{Sim}).  Figure \ref{Sim} shows that the
upper spin-subband de-populates more quickly for $B \parallel
[\bar233]$ than for $B
\parallel [01\bar1]$, which can be traced back to the
strong anisotropy of the 2DHS band structure in our system
\cite{SpinSp}. This is consistent with the experimental
observation that $B^*$ is smaller for $B \parallel [\bar233]$.
Also, the $B$ at which the subband completely de-populates changes
with $p$ in much the same way as $B^*$ does.  However, $B^*$ is
significantly smaller than the field at which the calculations
show the upper spin-subband to reach zero density. We believe that
the spin-subband de-population occurs at a lower $B$ than the band
calculations predict because hole-hole interaction enhances the
effective mass $m^*$ and effective $g$-factor $g^*$ in a dilute 2D
system like ours. The average hole spacing in units of effective
Bohr radius, $r_s$, for our experiment ranges from $r_s = 6.9$ to
15.6 for $p = 6.6 \times 10^{10}$ cm$^{-2}$ to $1.3 \times
10^{10}$ cm$^{-2}$ \cite{Rs}. Okamoto {\it et al.}
\cite{Okamoto99}, conclude that for Si 2D electrons with $r_s$ in
this range, $g^*m^*$ is enhanced by a factor of 2.7 to 5.5.
Assuming similar enhancement in our samples, we would expect a
reduction by the same factor of $B$ required to de-populate a
subband. Using these numbers to adjust the de-population field
given by the band calculations would put it near $B^*$, strongly
suggesting that the observed MR features are due to spin-subband
de-population.

Further evidence linking the MR features to spin-subband
de-population is provided by our data at constant $p$ with
changing $E_{\perp}$. The degree of asymmetry in the potential
that confines the carriers to 2D controls the spin-splitting, and
plays an important role in the magnitude of the $B = 0$
temperature-dependence of the resistivity
\cite{Papadakis99,Papadakis99c}. For the data in Figs. \ref{intro}
and \ref{density}, $E_{\perp}$ is kept within 1 kV/cm of 5 kV/cm.
This $E_{\perp}$ is included in the calculations plotted in Fig.
\ref{Sim}. Measurements at a constant $p = 3.9 \times 10^{10}$
cm$^{-2}$ as $E_{\perp}$ is increased from 4.5 kV/cm to 12.5 kV/cm
reveal that $B^*$ shifts to higher $B$ by about 2 T. This
observation is in agreement with the spin-subband de-population
calculations done at fixed $p$ for varying $E_{\perp}$.

At higher in-plane $B$, beyond the MR features around $B^*$, the
data in Fig. \ref{density} are qualitatively similar.  The traces
for $B \perp I$ have greater slope than the corresponding traces
with $B
\parallel I$, regardless of crystal axes.  In this regime the
magnetic confinement can become comparable to the electric
confinement, and the effects due to the finite-thickness of the
2DHS may be dominant. Indeed, Ref. \cite{DasSarmaC99} predicts
that MR with in-plane $B$ should be significantly larger for $B
\perp I$ than for $B \parallel I$, in agreement with our highest
$B$ data. The data in which $E_{\perp}$ is changed at constant $p$
support this interpretation as well. As $E_{\perp}$ is increased,
the confining potential becomes narrower, and the thickness of the
2DHS decreases. This should increase the $B$ required for
finite-thickness effects to become important, and the data show
that the MR anisotropy at $B = 16$ T is smaller for larger
$E_{\perp}$.

\begin{figure}[tb]
\epsfxsize=3.25in
\epsfbox{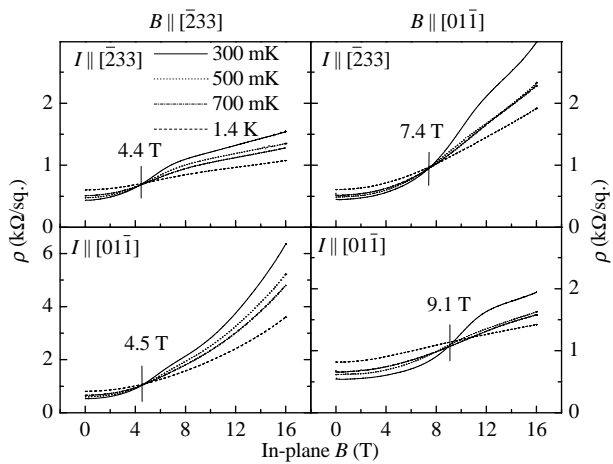}
\caption{Magnetoresistance data at various temperatures, for
density $p = 3.9 \times 10^{10}$ cm$^{-2}$, for the four relative
orientations of $B$, $I$, and crystal axes.  The fields $B_T$ at
which the resistivity is nearly $T$-independent are indicated by
vertical marks.} \label{temperature}
\end{figure}

We now turn to the $T$-dependence of MR to investigate the
metallic phase in our 2D system. Figure \ref{temperature} shows
the $T$-dependence of MR at $p = 3.9 \times 10^{10}$ cm$^{-2}$,
for the four measured relative orientations of $B$, $I$, and
crystal axes.  For each panel, the traces exhibit a nearly
$T$-independent magnetic field $B_T$ which occurs near the trace's
first inflection point.  This is consistent with the data of Ref.
\cite{Yoon99b}.  For $B < B_T$, the data show metallic behavior,
and for $B > B_T$, insulating behavior. $B_T$ is different in each
panel and, similar to $B^*$, it changes much more for a rotation
of the crystal axes relative to $B$ than it does for a rotation of
$I$ relative to $B$.  Our experiments indicate that $B^*$ and
$B_T$ depend very similarly on the parameters of our systems ($p$,
$E_{\perp}$, direction of $B$).  Our observation, which is in
agreement with the in-plane MR data of Ref. \cite{Okamoto99},
strongly suggests that the metallic behavior is linked to the
presence of two populated spin-subbands
\cite{Papadakis99,Papadakis99c,Murzin98,Yaish99}.

In the data of Ref. \cite{Okamoto99}, and likely in ours, the
spin-subband de-population is linked to $B^*$ which is somewhat
larger than $B_T$, so it appears that the metallic behavior
changes to insulating before the upper spin-subband is fully
de-populated. This may be because the low-density spin-subband
stops playing a role in transport when its mobility $\mu$ becomes
sufficiently low, before it is fully de-populated.

Finally, we note that Das Sarma and Hwang have recently reported
calculations aiming to explain the $T$-dependence of the
resistivity \cite{DasSarma99} and the in-plane MR
\cite{DasSarmaC99} of 2D systems that exhibit metallic behavior at
finite $T$.  Their calculations, which include only charged
impurity scattering and the orbital motion, qualitatively
reproduce some of the experimental data.  We wish to point out
that our results reveal the importance of the spin degree of
freedom, and suggest that for an understanding of the experimental
data it is important to also consider a scattering mechanism
involving the spin-subbands, perhaps intersubband scattering
\cite{Murzin98,Yaish99}. Also important for (311)A GaAs 2D holes
is the inclusion of interface roughness scattering:  both the
$T$-dependence of $\rho$ at $B = 0$ (Fig. \ref{aniso}b), as well
as the in-plane MR data (Figs. \ref{intro} and \ref{density}),
depend on the direction of the current in the crystal.

To summarize, our data reveal a surprising anisotropy of the
in-plane magnetoresistance and its temperature dependence for GaAs
(311)A 2D holes.  The results show that the rate of the upper
spin-subband's de-population with in-plane $B$ critically depends
on the relative orientation of $B$ and the crystal axes.  This
points to the anisotropic nature of the $g$-factor and the
spin-subband structure of GaAs (311)A 2D holes.  Furthermore, we
observe that the $B$ = 0 metallic behavior turns into insulating
near $B$ at which the upper spin-subband de-populates.  This
observation, in agreement with the data for 2D electrons in Si
\cite{Okamoto99}, suggests that two spin-subbands are necessary
for the expression of metallic behavior.  These results also
complement those of previous experiments
\cite{Papadakis99,Papadakis99c,Murzin98,Yaish99} which revealed
that the presence of two spin-subbands with different populations
appears to be linked to the metallic behavior.

This work was supported by the NSF and ARO.  We thank M. Hofmann
for stimulating discussions.


\begin{thebibliography}{10}
\bibitem{SpinSp}
For $B = 0$ the upper spin-subband density is smaller than half
the $p$ because
  of the inversion-asymmetry-induced spin splitting of the subband states [9].
\end{thebibliography}

\end{document}